\documentclass[useAMS,usenatbib,usegraphicx]{mn2e}

\usepackage{ulem}

\title{Shock-induced star formation in a model of the Mice}

\author[J.E. Barnes]{Joshua E. Barnes\thanks{E-mail:
barnes@ifa.hawaii.edu}\\
Institute for Astronomy, 2680 Woodlawn Drive, Honolulu, HI 96822, USA}

\begin{document}

\pagerange{\pageref{firstpage}--\pageref{lastpage}} \pubyear{2003}

\maketitle

\label{firstpage}

\begin{abstract}
Star formation plays an important role in the fate of interacting
galaxies.  To date, most galactic simulations including star formation
have used a density-dependent star formation rule designed to
approximate a Schmidt law.  Here, I present a new star formation rule
which is governed by the local rate of energy dissipation in shocks.
The new and old rules are compared using self-consistent simulations
of NGC\,4676; shock-induced star formation provides a better match to
the observations of this system.
\end{abstract}


\section{INTRODUCTION}

Numerical simulations of galaxy formation and interactions often
include rules for star formation.  In many cases, the local rate of
star formation, $\dot{\rho}_{*}$, is related to the local gas density,
$\rho_{\rm g}$, by a power law:
\begin{equation}
  \dot{\rho}_{*} = C_{*} \, \rho_{\rm g}^n \, ,
  \label{sfr-power-law}
\end{equation}
where $C_{*}$ is a constant.  This prescription has been justified
both empirically and theoretically.  From an empirical perspective,
(\ref{sfr-power-law}) resembles the ``Schmidt law'' \citep{S59}, which
relates star formation per unit surface area, $\dot{\Sigma}_{*}$ to
gas surface density, $\Sigma_{\rm g}$; a recent determination of the
Schmidt law \citep{K98} is
\begin{equation}
  \dot{\Sigma}_{*} = (2.5 \pm 0.7) \times 10^{-4} \,
    \frac{{\rm M}_{\sun}}{{\rm yr} \, {\rm kpc}^{2}} \,
      \left( \frac{\Sigma_{\rm g}}{1 {\rm M}_{\sun} {\rm pc}^{-2}} \right)
	^{1.4 \pm 0.15} \, .
  \label{schmidt-kennicutt-law}
\end{equation}
\citet{MRB91} adopted a rule equivalent to (\ref{sfr-power-law}) with
$n = 2$ as an approximation to a Schmidt law with index $\sim 1.8$,
while \citet{MH94a} took $n = 1.5$, and presented numerical tests
showing this gave a reasonable match to a Schmidt law with index $\sim
1.5$.  A more theoretical approach, adopted by \citet{K92} and
\citet{S00}, sets $\dot{\rho}_{*} = \rho_{\rm g} / t_{*}$, where
$t_{*}$, the time-scale for star formation, is basically proportional
to the local collapse time of the gas, $(G \rho_{\rm g})^{-1/2}$.
This yields (\ref{sfr-power-law}) with $n = 1.5$ and $C_{*} = G^{1/2}
C'_{*}$, where $C'_{*}$ is a dimensionless constant.

Despite the apparent convergence of observation and theory on the
index $n = 1.5$, it's unlikely that (\ref{sfr-power-law}) really
captures the process of star formation.  For one thing, only about
$1$~per cent of the gas actually forms stars per collapse time $(G
\rho_{\rm g})^{-1/2}$; in other words, consistency with the
observations implies that $C'_{*} \simeq 10^{-2}$.  Current theories
of star formation don't offer any straightforward way to calculate
this quantity.  Gravitational collapse is evidently not the limiting
factor which sets the rate of star formation; ``feedback'' processes
are important in determining the fraction of available interstellar
material which ultimately becomes stars.  While several groups have
now devised simulations including star formation regulated by various
forms of feedback (e.g.~\citealt{GI97}; \citealt{S00}), this approach
still has some way to go.

Moreover, models based on (\ref{sfr-power-law}) don't reproduce the
large-scale star formation seen in many interacting galaxies.
\citet{MRB92} found that most of the star formation was confined to
the central regions of their model galaxies, while \citet{MBR93} noted
that (\ref{sfr-power-law}) underestimated the rate of star formation
in regions where interstellar gas exhibits large velocity dispersions
and gradients, or where the gas appears to be undergoing strong
shocks.  \citet{MH94b}, \citet{MH96}, and \citet{S00} have modeled the
central bursts of star formation seen in ultraluminous infrared
galaxies \citep[][and references therein]{SM96}.  But nuclear
starbursts, while a necessary stage in the transformation of merger
remnants into elliptical galaxies \citep[e.g.][]{KS92}, may not be
sufficient to accomplish this transformation.  For example,
merger-induced starbursts create massive young star clusters
\citep[e.g.][]{WS95} which may subsequently evolve into globular
clusters, but these clusters will be confined to the nuclei of
remnants {\it unless\/} the starbursts are spatially extended.

In view of these considerations, it's worth examining alternatives to
(\ref{sfr-power-law}).  One long standing idea is that collisions of
molecular clouds trigger of star formation (e.g.~\citealt{SSC86}).
This has been implemented in ``sticky particle'' schemes which model
the interstellar medium as a collection of discrete clouds undergoing
inelastic collisions (e.g.~\citealt{OK90}; \citealt{N91}).  However,
molecular clouds have relatively long mean free paths, and direct
collisions between clouds at velocities of $10^2 {\rm\,km/s}$ or more
may result in disruption rather than star formation.  \citet{JS92}
proposed a model of shock-induced star formation in interacting
galaxies; specifically, they suggested that fast collisions between
extended H{\small{I}} clouds create a high-pressure medium which
compresses pre-existing molecular clouds and thereby induces bursts of
star formation.  High-pressure regions, and especially large-scale
shocks in colliding galaxies, may favor the formation of bound star
clusters \citep{EE97}.

Recent observations at optical, infrared, and radio wavelengths
continue to reveal large-scale star formation in interacting galaxies,
including NGC\,4038/9, Arp\,299, and NGC\,4676 (\citealt{WS95};
\citealt{V+96}; \citealt{M+98}; \citealt{HY99}; \citealt{AH+00};
\citealt{X+00}; \citealt{dG+03}).  Several of these studies explicitly
invoke shock-induced star formation in discussing the observational
data.  In this paper I focus on NGC\,4676, a strongly interacting pair
of disk galaxies with long tidal tails (\citealt{TT72}, hereafter
TT72; \citealt{T77}), as a test-case for models of shock-induced and
density-dependent star formation.  \S~2 describes the star formation
algorithms.  \S~3 presents simulations of NGC\,4676, and contrasts the
results of shock-induced and density-dependent star formation rules.
Conclusions are given in \S~4.

\section{SIMULATING STAR FORMATION}

Within the framework of ``Smoothed Particle Hydrodynamics'' or SPH
(\citealt{L77}; \citealt{GM77}; \citealt{M92} and references therein),
the interstellar material can be modeled as an isothermal gas with a
representative temperature $T \sim 10^4 {\rm\,K}$.  This approach is
less complex than others which include radiative cooling
\citep[e.g.][]{BH96}, heating by massive stars \citep{GI97}, or other
feedback effects \citep{S00}.  However, previous studies indicate that
simple isothermal models capture much of the relevant dynamical
behavior of interstellar gas in interacting galaxies, including
large-scale shocks, nuclear inflows, and the formation of extended gas
disks in merger remnants (\citealt{BH96}; \citealt{MH96};
\citealt{B02}).

With the gas temperature fixed, there remain two interesting local
quantities which are easily evaluated in an SPH simulation: the gas
density, $\rho_{\rm g}$, and the rate of mechanical heating due to
shocks and $PdV$ work, $\dot{u}$.  (An isothermal model assumes that
radiative processes locally balance mechanical heating or cooling.)
In galactic simulations, the gas temperature is much lower than the
virial temperature, so most shocks are highly supersonic and shock
heating exceeds $PdV$ work by about two orders of magnitude.  Thus
shocks can be identified as regions with $\dot{u}$ significantly
larger than zero; this criterion serves to locate shock fronts
\citep{B02}.

Making use of both $\rho_{\rm g}$ and $\dot{u}$, I consider a general
star-formation prescription of the form
\begin{equation}
  \dot{\rho}_{*} = C_{*} \, \rho_{\rm g}^n \, {\rm MAX}(\dot{u},0)^m \, ,
  \label{sfr-two-power-law}
\end{equation}
where the MAX function, which returns the larger of its two arguments,
is used to handle situations in which $\dot{u} < 0$.  This
prescription is implemented in a probabilistic fashion; the chance of
gas particle $i$ undergoing star formation in a time interval $\Delta
t$ is
\begin{equation}
  p_{i} = C_{*} \, \rho_{i}^{(n-1)} \,
            {\rm MAX}(\dot{u}_{i},0)^m \, \Delta t \, ,
  \label{sf-prob-law}
\end{equation}
where $\rho_{i}$ and $\dot{u}_{i}$, the gas density and heating rate
for particle $i$, are defined in Appendix~A.  At each time-step,
$p_{i}$ is computed for each gas particle and compared with a random
number $X$ freshly drawn from a uniform distribution between $0$ and
$1$.  If $p_{i} > X$ then the gas particle is instantly and completely
converted to a star particle, which subsequently evolves
collisionlessly.  For later analysis, each star particle formed during
a simulation is tagged with the value of the time $t$ at its moment of
birth.

While (\ref{sfr-two-power-law}) can encompass a wide range of rules,
this paper focuses on two limiting cases.  First, setting $n > 1$ and
$m = 0$ yields {\it density-dependent\/} star formation, in which
$p_{i} \propto \rho_{i}^{(n-1)}$ with {\it no\/} dependence on
$\dot{u}$, and regions of higher density produce more stars.  Second,
setting $n = 1$ and $m > 0$ yields {\it shock-induced\/} star
formation, in which $p_{i} \propto {\rm MAX}(\dot{u},0)^m$; star
formation is strictly limited to regions with $\dot{u} > 0$, which
typically correspond to shocks.  More general rules with both $n > 1$
and $m > 0$ are beyond the scope of this paper, but it may be possible
to anticipate their behavior using experience gained with the simple
cases described here.

Simulations of shock-induced star formation are subject to resolution
effects.  The width of a shock region in SPH is $\sim 3$ times the
local smoothing scale $h_i$ \citep{M92} (in these experiments, $h_i$
is set to enclose $N_{\rm smooth} = 40$ gas particles within a radius
$2 h_i$ of each gas particle $i$; see Appendix~B).  Consider a
situation in which a gas particle encounters a shock of some fixed
strength.  The time that the particle spends in the shock scales as
$h_i$, while the total amount of energy it dissipates is independent
of $h_i$, so $\dot{u}_i \propto h_i^{-1}$.  Thus the net probability
that the gas particle is transformed into a star particle while
passing through the shock is proportional to $h_i^{(1 - m)}$.  The
specific choice $m = 1$ should produce results which are approximately
independent of the spatial resolution.  For other choices for $m$ the
star formation rate will depend on resolution; to obtain roughly
equivalent results in simulations with different spatial resolutions,
it will be necessary to adjust the parameter $C_{*}$.

Resolution effects also arise in simulations with density-dependent
star formation rules.  Such simulations do a good job of reproducing
the observed relationship (\ref{schmidt-kennicutt-law}) between gas
density and star formation rate (\citealt{MH94a}; \citealt{S00}).  But
current SPH simulations generally don't resolve the vertical structure
of gas in galactic disks, so this success may be due, in part, to
compensating errors at different distances from the disk midplane.
More critically, although shocks with Mach numbers of $M \simeq 10$
to~$30$ will boost the density of an isothermal gas by factors of $M^2
\simeq 10^2$ to~$10^3$ \citep{S92}, these high-density regions are not
resolved in existing SPH calculations, and the localized bursts of
star formation which should result have not been seen in the numerical
models.  In principle, simulations using density-dependent rules can
exhibit shock-induced star formation, but the computing power required
to achieve the necessary resolution may not yet be available.

\begin{figure*}
\begin{center}
\includegraphics[width=177mm]{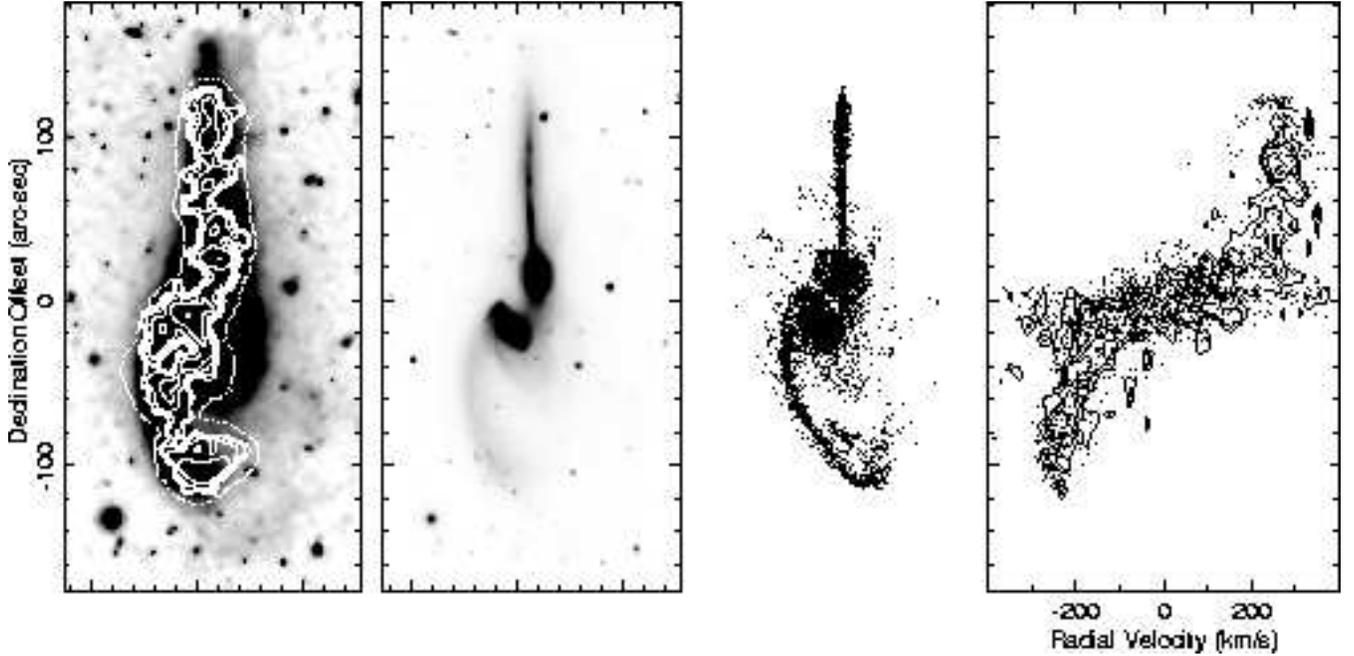}
\end{center}
\caption{The Mice, NGC\,4676, and the model.  North is up, west is
right.  The north-most hulk is NGC\,4676a, while its companion to the
southeast is NGC\,4676b.  Far left: deep optical image and
H{\small{I}} contours \citep{HvG96}.  Left: optical image, stretched
to show the bodies and tails.  Right: self-consistent N-body model.
Far right: declination versus line-of-sight velocity plane; contours
are H{\small{I}} data, points are N-body model.}
\label{fig01}
\end{figure*}

\section{MODELING THE MICE}

On account of their rounded bodies and long tails, the two galaxies
making up NGC\,4676 were dubbed the ``Playing Mice'' by \citet{V58}.
The far left panel of Fig.~\ref{fig01} shows a deep optical image,
overlain by contours of H{\small{I}} \citep{HvG96}, while the adjacent
panel shows a shallower image.  The individual galaxies are
NGC\,4676a, to the north, and NGC\,4676b, to the southeast.  The
length and straightness of the bright northern tail imply that we are
viewing the aftermath of a nearly-direct passage roughly edge-on to
the orbit and disk planes, while the curve of the fainter southern
tail is consistent with a more face-on view of an inclined disk.  The
fairly equal lengths of the two tails suggests that the galaxies
involved had roughly equal masses.  Along with long-slit data
indicating that both ``hulks'' rotate with north receding
(\citealt{BB61}; \citealt{TST72}), these considerations led TT72 to a
simple model for this system.  Their model required that the northern
hulk recede faster than the southern one, a prediction soon confirmed
with H$\alpha$ spectroscopy by \citet{S74}.  Most kinematic studies,
whether based on H$\alpha$ spectroscopy \citep{MBR93}, H{\small{I}}
interferometry \citep{HvG96}, or CO interferometry \citep{YH01}, yield
results consistent with \citeauthor{S74}'s.  On the other hand,
H$\alpha$ spectroscopy by \citet{SR98} indicated a remarkably large
difference of $\sim 300 {\rm\,km\,s^{-1}}$ between the nucleus of
NGC\,4676a and its associated tail, and a large velocity dispersion
within the tail itself; it's tempting to interpret this as measurement
error, but \citet{TST72} reported somewhat similar results, so the
situation is not entirely clear.  Nonetheless, H{\small{I}}
interferometry appears the best choice for tracing the large-scale
dynamics of the system, inasmuch as it provides a full map of the
velocity field, and neutral hydrogen is unlikely to be disturbed by
star formation.

\subsection{Dynamical model}

The model of NGC\,4676 used in this paper is based on work with
J.~Hibbard (Hibbard \& Barnes, in preparation; see also
\citealt{B98}).  In view of some concerns noted below, this model is
somewhat preliminary, but it seems to describe the overall evolution
of the system reasonably well.

The initial conditions for the individual galaxies each have three
components: a bulge with a shallow cusp \citep{H90}, an exponential
disk with constant scale height (\citealt{F70}; \citealt{S42}), and a
dark halo with a constant-density core, based on a ``gamma model''
(\citealt{D93}; \citealt{T+94}).  Density profiles for these
components are
\begin{equation}
  \begin{array}{@{}l}
    \rho_{\rm bulge} \propto
      r^{-1} (r + a_{\rm bulge})^{-3} \,, \\
    \rho_{\rm disk} \propto
      \exp(-R/R_{\rm disk}) \, {\rm sech}^2(z/z_{\rm disk}) \,, \\
    \rho_{\rm halo} \propto
      (r + a_{\rm halo})^{-4} \,,
  \end{array}
\end{equation}
where $r$ is the spherical radius, $R$ is the cylindrical radius in
the disk plane, and $z$ is the distance from the disk plane.
Numerical results are presented using a system of units with $G = 1$.
In these units, the components have masses $M_{\rm bulge} = 1/16$,
$M_{\rm disk} = 3/16$, $M_{\rm halo} = 1$, and length scales $a_{\rm
bulge} = 0.04168$, $R_{\rm disk} = 1/12$, $z_{\rm disk} = 0.005$, and
$a_{\rm halo} = 0.1$.  With these parameters, the disk has a rotation
period of $\sim 1.0$ time units at $R = 3 R_{\rm disk}$.  Scale
factors relating simulation units to real physical quantities will be
determined as part of the model-matching process.

Beginning with TT72's model, we ran experiments systematically varying
the relative orbit and orientation of the two identical galaxies.  The
following parameters produced the reasonably good fit to NGC\,4676
shown on the right and far right of Fig.~\ref{fig01}.  First, the
initial orbit was parabolic, with a pericentric separation $r_{\rm p}
= 0.25 = 3 R_{\rm disk}$.  Second, the disks of NGC\,4676a and
NGC\,4676b were initially inclined to the orbital plane by $i_{\rm a}
= 25^\circ$ and $i_{\rm b} = 40^\circ$.  Third, the initial angle
between each disk's line of nodes and the direction of separation at
pericenter was $\omega_{\rm a} = -30^\circ$ and $\omega_{\rm b} =
60^\circ$.  Fourth, the system was evolved $1$ time unit past
pericenter.  We used an interactive display program to rotate the
model to best match the observed morphology and kinematics NGC\,4676;
further details of the matching process will appear in a subsequent
paper.

Comparing the model with the observations, the main discrepancy is the
position angle of NGC\,4676b's bar; optical images show the bar runs
roughly NNE--SSW, while the bar in the model is roughly N--S or even
NNW--SSE.  The pattern speed of a bar depends on the details of the
rotation curve, so this discrepancy might be resolved by adjusting the
mass model used for NGC\,4676b.  It's also possible that NGC\,4676b's
bar predates its encounter with NGC\,4676a, in which case the bar's
current position angle would depend on its phase at pericenter
\citep{GCA90}.  However, the hypothesis that NGC\,4676b's bar was
induced by the tidal interaction is more parsimonious and has not yet
been tested by experiments using a range of mass models.

The time yielding the best match between the model and NGC\,4676
proved somewhat ambiguous; by adjusting the viewing angle and scaling
factors, acceptable matches to the tail morphology and kinematics were
obtained for $t \simeq 0.75$~to~$1.0$ time units after pericenter.  At
$t = 0.875$, a good match was obtained by equating one length unit to
$80'' = 35.7 {\rm\,kpc}$, and one velocity unit to $180
{\rm\,km\,s^{-1}} = 0.184 {\rm\,kpc\,Myr^{-1}}$.  This scaling assumes
that NGC\,4676's distance is $92 {\rm\,Mpc}$, consistent with $H_0 =
72 {\rm\,km\,s^{-1}\,Mpc^{-1}}$ and NGC\,4676's systemic velocity of
$6600 {\rm\,km\,s^{-1}}$ \citep{HvG96}.  One time unit then works out
to $194 {\rm\,Myr}$, and pericenter occurred $\sim 170 {\rm\,Myr}$
ago.  Finally, one mass unit is $2.69 \times 10^{11} {\rm\,M_\odot}$.

Given this scaling, the initial parameters of the model galaxies can
be computed.  For example, the initial disks have reasonable scale
lengths $R_{\rm disk} \simeq 3.0 {\rm\,kpc}$ and masses $M_{\rm disk}
\simeq 5.0 \times 10^{10} {\rm\,M_\odot}$; on the other hand, they
have maximum circular velocities of $\sim 290 {\rm\,km\,s^{-1}}$,
which seem rather high and might be repaired by increasing the halo
scale radius $a_{\rm halo}$.

Since TT72's work, NGC\,4676 has been modeled by several other groups.
\citet{MBR93} essentially produced a self-consistent clone of TT72's
model, even adopting a short-period elliptical orbit like the one
TT72's test-particle methodology dictated; they favored an earlier
viewing time which matched the position angle of NGC\,4676b's bar but
produced tails significantly shorter than those observed.
\citet{GS93} used \citeauthor{S74}'s (\citeyear{S74}) velocity data to
construct a self-consistent model with a parabolic initial orbit.  In
most respects, their model is very similar to the one described here,
although we favor a closer passage to increase the heft of the tidal
tails.  \citet{SR98} produced a test-particle model, restricted to
NGC\,4676a, which matched the general trend of their velocity data but
not the large velocity dispersion they reported within the northern
tail.  Their model requires relatively massive dark halos to account
for these high tail velocities.

\subsection{Interaction-induced star formation}

\begin{figure}
\begin{center}
\includegraphics[width=84mm]{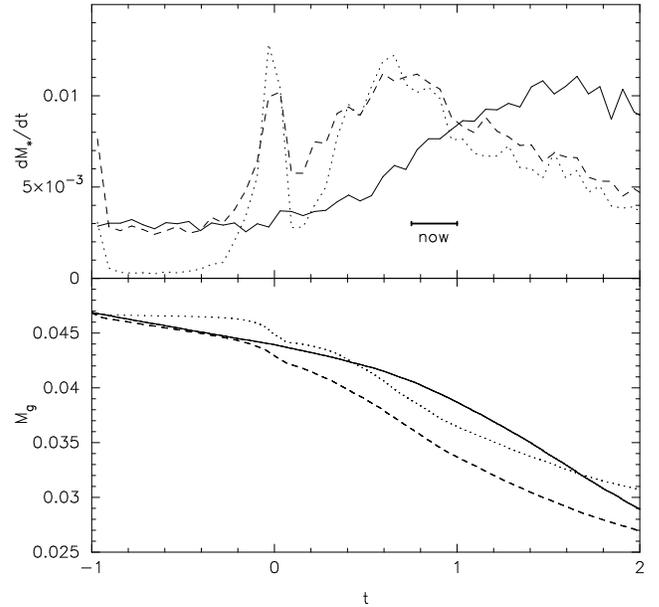}
\end{center}
\caption{Global star formation in simulations of NGC\,4676.  The top
panel shows $\dot{M}_{*}$, the rate of star formation, while the
bottom panel shows the mass of gas remaining; both quantities are
given in simulation units ($G = 1$).  Pericenter occurred at $t = 0$;
the horizontal bar labeled ``now'' indicates the range of times
matching the morphology and kinematics of NGC\,4676.  Solid line:
density-dependent star formation with $n = 1.5$, $m = 0$, $C_{*} =
0.025$.  Broken lines: shock-induced star formation; dashed is $n =
1$, $m = 0.5$, $C_{*} = 0.5$, dotted is $n = 1$, $m = 1$, $C_{*} =
0.25$.  Note that the shock-induced models produce strong bursts of
star formation at pericenter ($t = 0$), while the density-dependent
model exhibits little or no immediate response.}
\label{fig02}
\end{figure}

Fig.~\ref{fig02} compares global star formation rates for three
different simulations of NGC\,4676.  To illustrate the range of
possibilities, I contrast one simulation with density-dependent star
formation (solid line: $n = 1.5$, $m = 0$, $C_{*} = 0.025$) and two
simulations with shock-induced star formation (dashed line: $n = 1$,
$m = 0.5$, $C_{*} = 0.5$; dotted line: $n = 1$, $m = 1$, $C_{*} =
0.25$).  All of these simulations were based on the initial orbit and
disk orientations described in \S~3.1; the gas was initially
distributed like the disk stars, and the gas mass in each disk was
$M_{\rm gas} = M_{\rm disk}/8 = 3/128$.  Further numerical details are
given in Appendix~B.

\begin{figure*}
\begin{center}
\includegraphics[width=177mm]{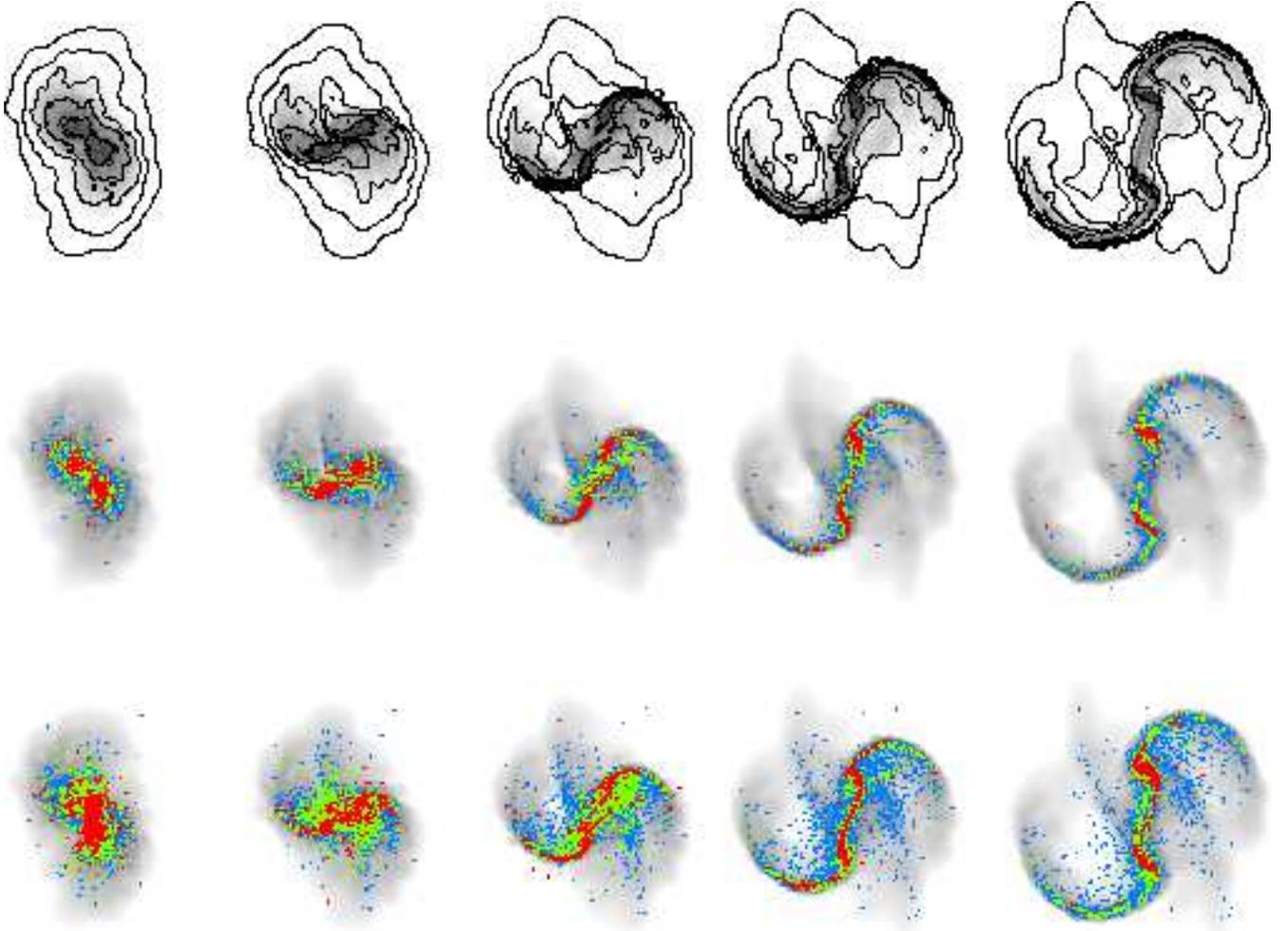}
\end{center}
\caption{Dynamical evolution and star formation in simulations of
NGC\,4676.  Top row: old stellar distribution (contours) and gas
density (halftone) at equally-spaced times between $t = 0$
(pericenter) and $0.5$.  Middle row: old stellar distribution
(halftone) and star formation (points) for the density-dependent model
($n = 1.5$, $m = 0$, $C_{*} = 0.025$).  Bottom row: old stellar
distribution (halftone) and star formation (points) for the
shock-induced model ($n = 1$, $m = 0.5$, $C_{*} = 0.5$).  Red points
have ages $\tau < 1/16$ time units, green have $\tau < 1/4$, and blue
have $\tau < 1$.  See
http://www.ifa.hawaii.edu/$\sim$barnes/research/interaction\_models/mice/
for animated versions of this and other figures.}
\label{fig03}
\end{figure*}

In simulations with density-dependent star formation, setting the
constant $C_{*} = 0.025$ yielded a baseline star formation rate per
disk of $\dot{M}_{*} \simeq 0.00136$ (see Appendix~C); at this rate,
an undisturbed disk would consume its gas on a time-scale of $M_{\rm
gas} / \dot{M}_{*} \simeq 17$ time units ($\sim 3.3 {\rm\,Gyr}$).  In
the simulation with shock-induced star formation and $m = 0.5$, a
similar baseline rate of $\dot{M}_{*} \simeq 0.00126$ was obtained by
setting $C_{*} = 0.5$; this gave reasonable behavior throughout the
encounter.  On the other hand, it was necessary to adopt a much lower
baseline rate of $\dot{M}_{*} \simeq 0.000175$ in the simulation of
shock-induced star formation with $m = 1$ to avoid using up most of
the gas in the earliest stages of the encounter; setting $C_{*} =
0.25$ insured that this case ultimately consumed about as much gas as
the other two.

As Fig.~\ref{fig02} shows, the density-dependent simulation predict
that the star formation rate remains fairly constant until some time
after the first passage.  Eventually, as the gas encountered shocks
and gravitationally interacted with the tidally induced bars in the
stellar disks, it began accumulating in the centers of the galaxies,
driving a fairly gradual increase in star formation.  The two
shock-induced simulations, on the other hand, produced sharp bursts of
star formation as the disks interpenetrated ($t \simeq 0$), followed
by brief lulls as the disks separated.  Soon thereafter, star
formation rates rose again as gas flows intersected in the perturbed
disks; this produced the broad peaks, centered on $t \simeq 0.6$,
accounting for most of the stars formed before $t = 2$.  As one might
expect, the simulation with $m = 1$ exhibited the most extreme form of
this behavior, with a higher peak at $t = 0$ and a deeper minimum
immediately thereafter.

These various star formation histories have interesting observational
implications.  According to the density-dependent simulation, star
formation in NGC\,4676 has just begun responding to the encounter, and
the peak star formation rate will occur several hundred Myr from now.
In contrast, the shock-induced models predict that we are observing
the system a short time after the broad peaks of star formation at $t
\simeq 0.6$.  This difference in star formation history is fairly
insensitive to the choice of $C_{*}$, which influences the overall
amplitude of star formation but does little to change the timing of
starbursts.

\begin{figure*}
\begin{center}
\includegraphics[width=177mm]{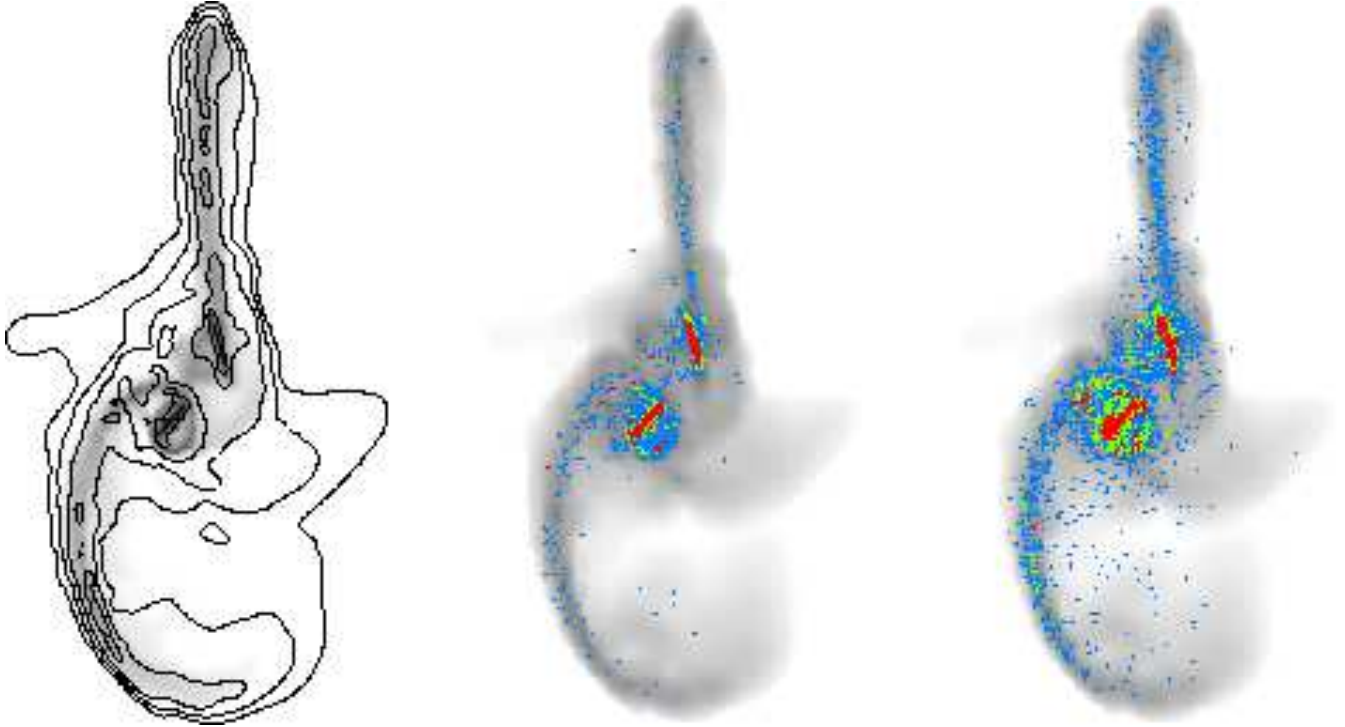}
\end{center}
\caption{Simulations matching the present configuration of NGC\,4676.
Left: old stellar distribution (contours) and gas density (halftone).
Middle: old stellar distribution (halftone) and star formation
(points) for the density-dependent model ($n = 1.5$, $m = 0$, $C_{*} =
0.025$).  Right: old stellar distribution (halftone) and star
formation (points) for the shock-induced model ($n = 1$, $m = 0.5$,
$C_{*} = 0.5$).  Colors indicate ages as in Fig.~\ref{fig03}.}
\label{fig04}
\end{figure*}

Fig.~\ref{fig03} contrasts the spatial distribution of star formation
in density-depend\-ent and shock-induced simulations.  The top row of images set
the stage by showing distributions of old disk stars and gas, taken
from a simulation without star formation.  The middle row shows
distributions of old disk stars and star particles formed according to
the density-dependent rule ($n = 1.5$, $m = 0$, $C_{*} = 0.025$);
colors indicate ages of star particles, with red being youngest and
blue oldest.  The bottom row presents results for a simulation with
shock-induced star formation ($n = 1$, $m = 0.5$, $C_{*} = 0.5$) using
the same color coding.  Results for the other shock-induced simulation
are not shown since they are very similar.

Comparison of the middle and bottom rows of Fig.~\ref{fig03} shows
that the star formation rule clearly influences the distribution of
newly-formed stars.  In the density-dependent model, star formation
was always strongly concentrated toward the central regions of the
disks.  Some star formation did occur outside the centers, but it was
relatively weak and short-lived.  In contrast, the shock-induced
models produced large, spatially-extended bursts of star formation as
the disks collided at $t = 0$, and relatively high levels of activity
persisted in the bridge and tail regions until $t \simeq 0.4$.

By the last time ($t = 0.5$) shown in Fig.~\ref{fig03}, products of
star formation are much more widely distributed in the shock-induced
simulations than in the density-dependent simulation.  Rather
interesting in this regard are the gas-poor features which extend to
roughly five o'clock from the upper galaxy and to eleven o'clock from
the lower galaxy.  These regions were swept free of gas during the
interpenetrating encounter of the two disks.  In the simulations with
shock-induced star formation, some of this gas was converted to stars
{\it before\/} losing much momentum; as a result, star formed in the
pericentric bursts are scattered throughout these gas-poor regions.
By contrast, these regions are bereft of both gas {\it and\/}
star-formation products in the density-dependent simulation.

Fig.~\ref{fig04} presents views which approximate the observed
morphology and kinematics of NGC\,4676.  These views are plotted at
time $t = 0.875$, but fairly similar results are found between $t =
0.75$ and $t = 1$.  The simulations shown here are the same ones shown
in the last figure, with the density-dependent case in the middle and
the shock-induced one on the right.  In both cases, most of the
ongoing star formation is now confined to central regions of the
galaxies.  However, products of star formation are found throughout
the tails and other tidal features, especially in the simulation with
shock-induced star formation.  These results invite comparison with
observations of past star formation in NGC\,4676.  \citet{S74} found
that the spectrum of NGC\,4676 is dominated by A stars, and attributed
this to ``rapid, widespread star formation that effectively ceased at
least $5 \times 10^7$ years ago''; this diagnosis is supported by more
recent observations (\citealt{MBR93}; \citealt{HvG96}; \citealt{SR98};
\citealt{dG+03}).  The shock-induced simulations produced just such
bursts at first passage ($t \simeq 0$), and the products of these
bursts are widely distributed throughout the tidal tails at later
times.  In contrast, the density-dependent simulation had far less
star formation outside the galactic centers, and the first passage did
not induce a prompt burst of extended star formation.

NGC\,4676 has regions of ongoing star formation which are evident in
H$\alpha$ images (\citealt{H95}, Fig.~AIII-1).  The central regions of
the two galaxies are quite prominent in H$\alpha$ (\citealt{MBR93});
central activity is seen in all the simulations.  Outside of the
centers of the galaxies, star-forming regions are observed in the
straight tail extending to the north of NGC\,4676a.  None of the
simulations presented here reproduce this phenomenon, although the
shock-induced models come closer inasmuch as they show enhanced star
formation in their tails until $t \simeq 0.4$; the density-dependent
model seems less promising in this regard since star formation in its
tails declines after first passage.  H$\alpha$ images also show
ongoing star formation at the base of the curved tail to northeast of
NGC\,4676b.  This region has a relatively high surface brightness and
is conspicuously blue in the ACS images (e.g.~\citealt{dG+03}).  It is
nicely reproduced in the shock-induced simulations; gas from the
curved tail falls back into its original disk, and the resulting shock
sustains a small, coherent patch of extra-nuclear star formation.

On the whole, the simulations with shock-induced rules qualitatively
match most of the recent large-scale star formation in NGC\,4676.
Eventually, this comparison can be made more precise by modeling the
spectrophotometric evolution of star particles, but such refinements
are beyond the scope of this paper.  The balance between central and
extended star formation can be quantified by measuring the mass in
stars formed since the start of the interaction.  Specifically, the
production of stars was integrated from $t = -0.25$, which is when
Fig.~\ref{fig02} shows the first increase in interaction-induced star
formation, until $t = 0.875$; particles then found at radii $r > 0.15$
from the centers of both galaxies were considered to be part of the
extended tidal features.  Table~\ref{starmass} list the results for
all three experiments.  At the epoch matching the present
configuration of NGC\,4676, the shock-induced models have formed about
twice as many stars as the density-dependent model; of course, this
ratio depends on the values chosen for $C_{*}$.  In the shock-induced
simulations, $28$~percent of the newly-formed stars are found within
the tidal features, while the corresponding figure for the
density-dependent simulation is only $13$~percent.  These percentages
should be insensitive to reasonable variations of $C_{*}$, and
preliminary experiments with other parameter values support this
expectation.

\begin{table}
\caption{Mass of stars formed since start of interaction, expressed as
a percentage of initial gas mass.}
\label{starmass}
\begin{tabular}{@{}rrrrr@{}}
  $n$ &   $m$ & extended & central & total \\
$1.5$ & $0.0$ &      1.4 &     9.2 &  10.6 \\
$1.0$ & $0.5$ &      5.9 &    14.7 &  20.6 \\
$1.0$ & $1.0$ &      5.2 &    13.6 &  18.8 \\
\end{tabular}
\end{table}

\subsection{Starbursts in mergers}

The future evolution of the NGC\,4676 system may seem a matter of
conjecture.  Nonetheless, it's worth asking when these galaxies are
likely to merge and if their merger will produce an ultraluminous IR
galaxy; even if we can't check the specific predictions for this
system, the results may provide insight into other merging galaxies.
I therefore present some predictions for the fate of NGC\,4676.

\begin{figure}
\begin{center}
\includegraphics[width=84mm]{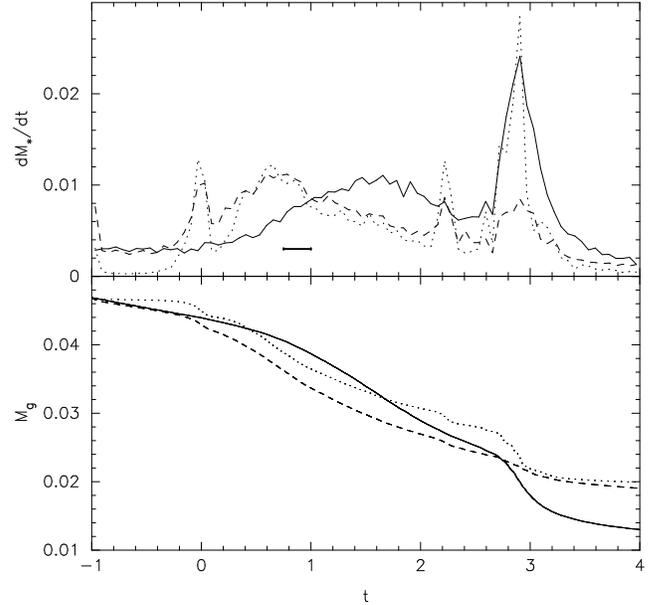}
\end{center}
\caption{Global star formation in simulations of NGC\,4676.  First
pericenter occurs at $t = 0$, second pericenter occurs at $t \simeq
2.2$, and the galaxies merge at $t \simeq 2.8$.  Line types are
identical to those in Fig.~\ref{fig02}.}
\label{fig05}
\end{figure}

Fig.~\ref{fig05} extends the comparison of global star formation rates
in NGC\,4676 some $600 {\rm\,Myr}$ into the future.  In these
simulations, star formation is modulated by two key events.  First, as
a result of orbital decay, the galaxies have a second and much closer
passage at $t \simeq 2.2$; this passage has little immediate effect in
the density-dependent model, but produces definite bursts in the
shock-induced models, especially in the one with the steeper $m = 1$
dependence on $\dot{u}$.  Second, the galaxies merge at $t \simeq
2.8$, producing dramatic bursts of star formation in the
density-dependent model and in the shock-induced model with $m = 1$,
but only a modest burst in the shock-induced model with $m = 0.5$.

In the shock-induced simulations, the bursts at the second passage
were produced by the same mechanism responsible for the comparable,
albeit somewhat larger, bursts at first pericenter: interpenetrating
encounters between two gassy disks.  While much of the gas originally
populating these disks was driven into the nuclei or converted into
stars, gas returning from the tidal tails settled back into the disks
(e.g. \citealt{B02}); thus, even after the fireworks attending their
first passage, the disks had enough extended gas to produce brief
starbursts as they came together for a second time.  It's worth noting
that such ``second-passage'' activity is apparently occurring in
Arp\,299, which exhibits both an extended starburst only $4
{\rm\,Myr}$ old (e.g.~\citealt{AH+00}) and a long tidal tail with an
age of $\sim 750 {\rm\,Myr}$ (\citealt{HY99}).

After the second passage, the orbits of the galaxies decayed very
rapidly, and a final, almost head-on encounter resulted in merger at
about $t = 2.8$.  As noted above, this event produced fairly intense
bursts of star formation in two of the three cases studied here.  In
the density-dependent simulation, this burst resulted from a sudden
increase in gas density as the nuclei finally coalesced.  Similar
results have been seen in other numerical studies using
density-dependent star formation; \citet{MH96} report bursts with peak
star formation rates as much as $10^2$ times the baseline.  The
present calculation yielded a more modest burst, peaking a factor of
$\sim 8$ above the baseline; the lower peak in the present simulation
may be due to consumption of gas at times $t \simeq 1$ to~$2$ while
the galaxies were still separate.  \citet{S00}, combining
density-dependent star formation with a sophisticated feedback scheme,
obtained final bursts comparable to the one shown here.  On the other
hand, \citet{MBR93} reported only a ``weak merger-induced'' starburst
in their model of NGC\,4676; while they blame this fizzle on the long
delay between first passage and merger in their dynamical model, it's
possible that the steeper $n = 2$ exponent of their star formation
prescription had some role in consuming almost all available gas
before the final encounter.

Up until this point, the two simulations with shock-induced star
formation have produced fairly similar results.  Nonetheless, as
Fig.~\ref{fig05} shows, their final bursts are quite different; the $m
= 0.5$ model yields a small burst with a maximum $\dot{M}_{*} \simeq
0.0085$, while the $m = 1$ model yields a sharply-peaked burst with a
maximum $\dot{M}_{*} \simeq 0.0285$.  In principle, this difference
could originate at earlier times -- for example, the $m = 0.5$ model
might yield a smaller final burst because it had already used up its
gas -- but a detailed examination of the simulations does not support
this idea.  At the start of their final encounters, both simulations
have comparable amounts of high-density nuclear gas, and in both
simulations these central regions collided almost head-on, producing
comparable amounts of dissipation.  Thus, the difference in burst
strengths reflects the different behavior of the $m = 0.5$ and $m = 1$
rules under nearly identical circumstances: in both cases, a moderate
amount of gas encountered a strong shock as the nuclei collided.  The
steeper dependence on $\dot{u}$ in the $m = 1$ case converted a large
fraction of the shocked gas into stars, while in the $m = 0.5$ case
the probability of star formation is less strongly enhanced and less
gas was consumed.  While the final bursts are not as dramatic as those
reported by \citet{MH96}, it's possible that further simulations of
shock-induced star formation with $m = 1$ could produce bursts large
enough to power luminous IR galaxies.

\begin{table}
\caption{Log median radii of burst populations.}
\label{burstradii}
\begin{tabular}{@{}rrrrrrr@{}}
  $n$ &   $m$ & B$_1$ & B$_2$ & B$_3$ & B$_4$ & all   \\
$1.5$ & $0.0$ &       & -1.57 &       & -2.44 & -1.78 \\
$1.0$ & $0.5$ & -0.55 & -1.10 & -1.34 & -1.91 & -1.23 \\
$1.0$ & $1.0$ & -0.49 & -1.10 & -1.09 & -1.85 & -1.30 \\
\end{tabular}
\end{table}

\begin{table}
\caption{Mass of burst populations, expressed as a percentage of
 initial gas mass.}
\label{burstmass}
\begin{tabular}{@{}rrrrrrr@{}}
$n$   & $m$   & B$_1$ & B$_2$ & B$_3$ & B$_4$ & all  \\
$1.5$ & $0.0$ &       &  29.9 &       &  16.6 & 72.1 \\
$1.0$ & $0.5$ &   5.1 &  14.1 &   3.2 &   4.9 & 59.5 \\
$1.0$ & $1.0$ &   4.7 &  13.9 &   4.1 &  11.3 & 57.6 \\
\end{tabular}
\end{table}

Tables~\ref{burstradii} and~\ref{burstmass} summarize properties of
the starbursts produced in these simulations.  Here a burst is defined
as all stars formed in a certain time interval; these intervals
bracket various peaks of $\dot{M}_{*}$ in Fig.~\ref{fig05}.  For the
shock-induced simulations, burst ``B$_1$'' corresponds to $-0.20 \le t
\le 0.12$, ``B$_2$'' to $0.28 \le t \le 0.93$, ``B$_3$'' to $2.10 \le
t \le 2.34$, and ``B$_4$'' to $2.69 \le t \le 3.00$.  For the
density-dependent simulation, ``B$_2$'' corresponds to $0.80 \le t \le
2.34$, and ``B$_4$'' to $2.69 \le t \le 3.15$.  In effect, bursts
B$_1$ and B$_3$ occur when the galaxies interpenetrate at first and
second passages, respectively, B$_2$ is the broad peak of star
formation in the individual galaxies after their first passages, and
B$_4$ is the final merger-induced starburst.  The last column in each
table, labeled ``all'', gives figures for all stars formed during
these simulations ($-1 \le t \le 4$).

The radial distribution of burst populations in the final ($t = 4$)
merger remnants is characterized in Table~\ref{burstradii}.  Here, the
center of each remnant ${\bf r}_{\rm cent}$ was defined by locating
the minimum of the potential well; next, the median of $|{\bf r}_i -
{\bf r}_{\rm cent}|$ for all particles $i$ in each burst population
was computed; common logarithms of these median radii are listed.  On
the whole, the burst populations in the shock-induced simulations are
$\sim 3$ times more extended than those in the density-dependent
simulation; this ratio applies not only to the stars produced by
bursts B$_2$ and B$_4$ but also to the entire population of stars
formed during the simulations.  The extended distribution of star
formation products in the shock-induced models may provide a better
match to galaxies like NGC\,7252, where \citet{S90} found spectra
dominated by A stars both within the nucleus and at a projected
distance of $7 {\rm\,kpc}$.  In addition, the final B$_4$ populations
in the shock-induced simulations, while still compact, may be extended
enough to moderate the central ``spikes'' of young stars that
\citet{MH94c} found in their simulated merger remnants.

Table~\ref{burstmass} lists masses of stars formed.  These masses are
given as percentages of the total initial gas mass, $2 M_{\rm gas} =
0.046875$.  The two shock-induced simulations, by design, formed
comparable amounts of stars, while the density-dependent simulation
ultimately formed somewhat more.  Burst B$_2$, triggered within the
individual galaxies after their first passage, yielded $41$~percent of
all stars formed in the density-dependent model, but only $24$~percent
in the shock-induced models.  The final burst B$_4$ accounts for
between $8.2$~percent and $23$~percent of the total star formation.

Earlier merger simulations including shock-induced star formation have
largely been based on ``discrete cloud'' models of the ISM, making a
direct comparison with the present results difficult.  \citet{N91}
presented cloud-collision rates for several mergers; assuming that
each cloud collision resulted in star formation, he predicted that
merging galaxies undergo ``repetitive starbursts'' as the relative
orbit of their nuclei decays.  The present calculations also produce
bursts of star formation associated with pericentric passages, but
orbit decay in these fully self-consistent calculations is much more
abrupt than in \citeauthor{N91}'s experiments, possibly because the
self-gravity of the gas is included.  Thus, while there is a good deal
of temporal structure in the star formation rate, the nearly-periodic
starbursts \citeauthor{N91} reported, which seem to reflect the
gradual decay of elongated orbits in a nearly-harmonic core, are not
seen here.  \citet{OK90} modeled the evolution of a spectrum of cloud
masses during a merger, and reported a dramatic increase in the rate
of disruptive cloud-cloud collisions within the inner $2 {\rm\,kpc}$
of their model galaxies.  Assuming that star formation is associated
with disruptive and glancing cloud-cloud collisions, they obtained
peak luminosities and luminosity-to-gas ratios typical of luminous IR
galaxies.

\section{CONCLUSIONS}

Density-dependent and shock-induced models of star formation yield
qualitatively and quantitatively different results.  These differences
arise despite the fact that the gas density $\rho_{\rm g}$ and the
dissipation rate $\dot{u}$ are not completely independent variables;
indeed, they can be explained in terms of the global relationship
between $\rho_{\rm g}$ and $\dot{u}$.  Dissipation is a necessary
precursor for any significant increase in central gas density, since
gas can only accumulate in galactic centers as a result of an
irreversible process.  The rate of increase in gas density closely
tracks the net energy radiated away in shocks (\citealt{BH96},
Fig.~6); globally, $\dot{u}$ and $d\rho_{\rm g}/dt$ are strongly
correlated.  Thus, models of shock-induced star formation can respond
promptly to external disturbances, while little activity occurs in
density-dependent models until $\rho_{\rm g}$ has had time to build
up.  The larger spatial extent of star formation in shock-induced
models is, in part, a corollary of the earlier onset of activity in
such models, since the gas is more widely distributed at earlier
times.

Density-dependent rules may be used to implement unified models of
star formation in normal and interacting galaxies.  The disks of
normal galaxies {\it and\/} the central regions of starburst galaxies
fit onto the same power law \citep{K98}.  Thus, by setting $C_{*}$ to
match the baseline rate of star formation in unperturbed disks,
\citet{MH96} were able to produce starbursts comparable to those
inferred in ultraluminous infrared galaxies \citep{SM96}; this
represents a real success for the model defined by
(\ref{sfr-power-law}) with $n = 1.5$.

But there is abundant evidence for large-scale star formation in
interacting galaxies.  This includes ongoing star formation in the
``overlap regions'' of systems like NGC\,4038/9 and Arp\,299, as well
as the H$\alpha$ emission from the tails of NGC\,4676.  It also
includes the A-star spectra seen in the tails of NGC\,4676 and
throughout the body of NGC\,7252.  Density-dependent star formation
can't easily explain these observations; interactions funnel gas from
the disks to the central regions of galaxies, thereby promoting rapid
star formation in galactic nuclei but reducing the supply of gas
needed for star formation elsewhere.  Moreover, violent relaxation
ceases long before binding energies are effectively randomized
(e.g.~\citealt{W79}; \citealt{B92}), so merging is ineffective at
transporting the products of star formation outward from nuclei to the
bodies of merger remnants.  Thus, density-dependent rules offer at
best an incomplete description of star formation in interacting
galaxies.

In simulations with shock-induced rules, the star formation rate
depends on dynamical circumstances.  Shocks in unperturbed disks are
associated with transient spiral patterns, and matching the baseline
level of star formation depends on reproducing the ``right'' level of
spiral structure.  This, in itself, is a tough problem, and it's not
clear that shock-induced models will soon yield a unified description
of star formation in normal {\it and\/} interacting galaxies.  Until
this becomes possible, unperturbed disks can't be used to set $C_{*}$,
so the amplitude of a shock-induced starburst can't be predicted {\it
a priori\/}.  Nonetheless, shock-induced star formation is an
important element of galactic collisions, and its implementation in
numerical simulations is a useful step toward increased realism.

These simulations have contrasted two limits of
(\ref{sfr-two-power-law}): density-dependent star formation, with $n >
1$ and $m = 0$, and shock-induced star formation, with $n = 1$ and $m
> 0$.  These cases were chosen as instructive examples, but
(\ref{sfr-two-power-law}) is general enough to accommodate additional
possibilities.  First, setting $n > 1$ and $m$ just slightly larger
than $0$ would yield a modified law in which star formation is
proportional to $\rho_{\rm g}^n$ but occurs {\it only\/} in regions
with $\dot{u} > 0$; this resembles the rule adopted by \citet{K92},
who basically took $\dot{\rho}_{*} \propto \rho_{\rm g}^{3/2}$ but
restricted star formation to regions with convergent flows.  Notice
that cases with $0 < m \ll 1$ are {\it not\/} continuous with the case
$m = 0$; in the latter, star formation is strictly independent of
$\dot{u}$.  Second, setting $n > 1$ and $m > 0$ would yield hybrid
rules in which star formation depends on both $\rho_{\rm g}$ and
$\dot{u}$.  The consequences of such rules can sometimes be inferred
from the limiting cases considered here.  For example, in the
simulations of NGC\,4676 with shock-induced star formation, the gas
involved in burst B$_3$ has densities $\sim 10^2$ times higher than
the gas involved in burst B$_1$; setting $n > 1$ would boost B$_3$
relative to B$_1$ by a factor of $\sim 10^{2 (n - 1)}$, assuming that
neither burst was limited by the supply of gas.

Perhaps most exciting are the new avenues for research created by an
alternative description of star formation in galaxy interactions:

\begin{itemize}

\item
In contrast to density-dependent star formation, which seems fairly
insensitive to most details of galactic encounters \citep{MH96},
different encounter geometries may yield a variety of shock-induced
star formation histories.  A survey of different encounters, along the
lines of other surveys without star formation (e.g.~\citealt{B02}),
might establish if widespread star formation at first passage is
generic or limited to a subset of close encounters, and also determine
if shock-induced star formation can account for ultraluminous infrared
galaxies.

\item
Shock-induced star formation makes definite predictions about the {\it
timing\/} of starbursts triggered by galactic encounters.  As already
mentioned, models including the photometric evolution of starburst
populations could sharpen the comparison between simulations and
observations of systems like NGC\,4676.  In addition, these models
might predict the ages of embedded young star clusters; such
predictions could be checked by multi-object spectroscopy of the
clusters.

\item
In the context of merger simulations, shock-induced star formation
also yields predictions for the spatial distribution and kinematics of
starburst populations.  It would be quite interesting to compare these
predictions with observations of the distribution and kinematics of
metal-rich globular clusters in elliptical galaxies, which may have
been formed in merger-induced starbursts (e.g.~\citealt{ZA93}).

\item
Encounters between gas-rich galaxies at redshifts of a few seem a
natural setting for shock-induced star formation.  While many high-$z$
galaxies have peculiar morphologies (e.g.~\citealt{vdB+96}), the
bridges and tails characteristic of low-$z$ encounters (TT72) are not
very evident.  Instead, the optical morphology of these objects may be
dominated by rest-frame UV from widespread starbursts; if so, the
photometric modeling approach described above could help interpret
existing and forthcoming observations of interacting galaxies at
intermediate and high redshifts.

\end{itemize}

\section*{ACKNOWLEDGMENTS}

I thank John Hibbard for permission to discuss our modeling of
NGC\,4676 in advance of publication, and for helpful discussions.  I
also thank Chris Mihos, Lars Hernquist, and especially Francois
Schweizer for comments on this paper, and the referee for a prompt and
helpful review.

\appendix

\section{SPH EQUATIONS}

I use an adaptive SPH code in which each gas particle $i$ is assigned
smoothing length $h_i$ which depends on local conditions.  The gas
density at particle $i$ is estimated using the symmetric form
suggested by \citet{HK89},
\begin{equation}
  \rho_i = \sum_j m_j \, \overline{W}_{ij}(r_{ij}) \, ,
  \label{sph-rho}
\end{equation}
where $m_j$ is the particle mass and $r_{ij} \equiv |{\bf r}_i - {\bf
r}_j|$.  Here, $\overline{W}_{ij}(r) \equiv \frac{1}{2}[W(r, h_i) +
W(r, h_j)]$, and $W(r, h)$ is the spline interpolation kernel
\citep{ML85}, which vanishes identically for $r > 2h$.

As discussed in \S~2, the code uses an isothermal equation of state,
so the internal energy $u_i$ of each gas particle is constant.  One
may nonetheless compute the mechanical heating rate $\dot{u}_i$
associated with the pressure and (artificial) viscous forces on each
particle $i$, and assume that this heating is {\it exactly\/}
balanced by radiative processes.  This requires only a modest number
of computations over and above those required to compute the forces
themselves, and allows a global check of energy conservation.  The
mechanical heating rate is estimated using
\begin{equation}
  \dot{u}_i = \sum_{j \ne i} m_j \,
    \left(\frac{P_i}{\rho_i^2} + \frac{\Pi_{ij}}{2}\right) \,
      ({\bf v}_i - {\bf v}_j) \cdot
        \frac{\partial}{\partial {\bf r}_i} \overline{W}_{ij}(r_{ij}) \, ,
  \label{sph-udot}
\end{equation}
where $P_i = c_{\rm s}^2 \rho_i$ is the pressure at particle $i$ for
an isothermal gas with sound speed $c_{\rm s}$.  Here, $\Pi_{ij}$
approximates a combination of bulk and von~Neuman-Richtmyer artificial
viscosity \citep{M92},
\begin{equation}
  \Pi_{ij} =
    \left\{\matrix{
      \frac{\displaystyle -\alpha c_{\rm s} \mu_{ij} + \beta \mu_{ij}^2}
	     {\displaystyle (\rho_i + \rho_j)/2} \, , & \mu_{ij} < 0 \, ; \cr
      0                                               & {\rm otherwise\/} \cr}
    \right.
  \label{sph-visc}
\end{equation}
where $\alpha$ and $\beta$ are parameters of order unity,
\begin{equation}
  \mu_{ij} = \frac{({\bf r}_i - {\bf r}_j) \cdot ({\bf v}_i - {\bf v}_j)}
                    {r_{ij}^2 / h_{ij} + \eta^2 h_{ij}} \, ,
\end{equation}
$h_{ij} \equiv {1 \over 2} (h_i + h_j)$, and $\eta$ is a parameter of
order $10^{-2}$.

\section{SIMULATION DETAILS}

The disk galaxies used in the star formation simulations had a total
of $87040$ particles each, with $24576$ for the gas, $21504$ for the
stellar disk, $8192$ for the bulge, and $32768$ for the halo.  The gas
was initially distributed like the stellar disk, and accounted for
$12.5$ per cent of the total disk mass.  In simulation units, the gas
had a fixed sound speed $c_{\rm s} = 0.0966$, mimicking a warm ISM
with $T \sim 10^4 {\rm\,K}$.

These calculations were run with a SPH code featuring adaptive
smoothing and individual particle timesteps.  The smoothing radius
$h_i$ of each gas particle was set so that a sphere of radius $2 h_i$
contains exactly $N_{\rm smooth} = 40$ gas particles.  A modified
Courant condition with $\mathcal{C} = 0.25$ was used to determine
individual particle timesteps, which in some cases were as small as
$\Delta t_i = 1/32768$ time units.  The artificial viscosity
parameters appearing in (\ref{sph-visc}) were $\alpha = 1.0$ and
$\beta = 2.0$.  Gravitational forces were calculated using a modified
tree code with Plummer smoothing \citep{A63} and a smoothing length of
$\epsilon = 0.0125$.  Collisionless particles were integrated with a
fixed time-step of $\Delta t = 1/512$ time units.  Including the
energy radiated (or absorbed) by the gas in order to maintain a
constant temperature, energy was conserved to about $0.5$~percent
through the final merger of the galaxy models.  Depending on the star
formation parameters, each simulation took between $126$~and $245
{\rm\,hr}$ on a $1 {\rm\,GHz}$ processor; the run without star
formation was the slowest since its gas attained the highest
densities.

In practice, the limited resolution of computer simulations introduces
effects which complicate the interpretation of numerical experiments.
Smoothing over a sphere of radius $2 h_i$ inevitably suppresses
structures on smaller scales.  The gas disks used in this paper
provide good examples; for the simulation parameters given above, only
half of the disk particles have smoothing radii $h_i$ less than the
vertical scale height $z_{\rm disk}$!  Clearly, the vertical structure
of these disks is not resolved -- and the same is true for most SPH
simulations of interacting galaxies published to date.  In this
circumstance, SPH underestimates $\rho_{\rm g}$ close to the midplane,
and overestimates $\rho_{\rm g}$ far from the midplane.

\section{ISOLATED DISK MODELS}

To better characterize the behavior of the galaxy models used in the
simulations of NGC\,4676, I ran some test calculations of isolated
disks.  These can be used to establish the baseline rate of
star formation in non-interacting disks, and to examine the radial
distribution of star formation for various choices of $n$ and $m$.

\begin{figure}
\begin{center}
\includegraphics[width=84mm]{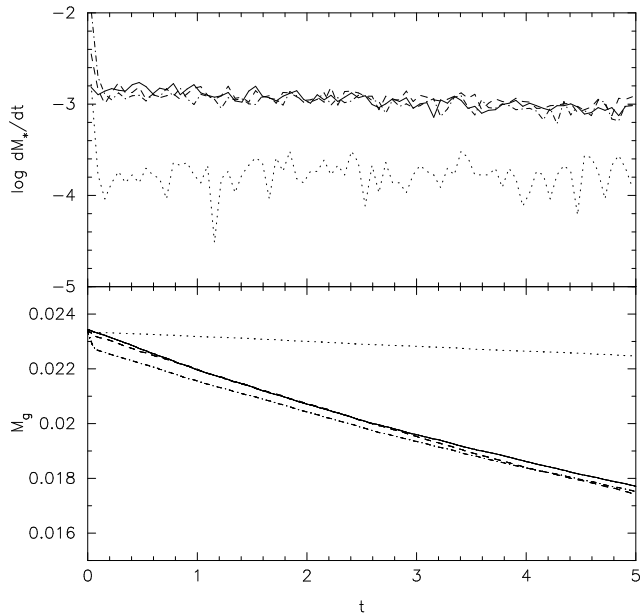}
\end{center}
\caption{Global star formation in isolated disk simulations.  Line
types are identical to those in Fig.~\ref{fig02}; in addition, the
dot-dashed line is $n = 1$, $m = 1$, $C_{*} = 2.0$.  Note that here
the top panel shows $\log \dot{M}_{*}$.  The simulations using
shock-induced star formation produce brief initial bursts, but settle
down promptly; no other features are seen in these plots.}
\label{fig06}
\end{figure}

Fig.~\ref{fig06} shows global star formation rates for isolated disk
models like the ones used in the NGC\,4676 simulations.  As in
Figs.~\ref{fig02} and~\ref{fig05}, the solid lines show results for
the density-dependent rule ($n = 1.5$, $m = 0$, $C_{*} = 0.025$) while
the broken lines show results for shock-induced rules (dashed line: $n
= 1$, $m = 0.5$, $C_{*} = 0.5$; dotted line: $n = 1$, $m = 1$, $C_{*}
= 0.25$).  Since the last of these yields a very low baseline rate, an
additional calculation was run with $C_{*}$ increased by a factor of
$8$ (dot-dashed line: $n = 1$, $m = 1$, $C_{*} = 2.0$).  In every
case, the isolated disk models proved stable for the duration of the
calculation.  Apart from the brief initial bursts in the shock-induced
models due to imperfect initialization of gas velocities, the only
trends seen are {\it slight\/} declines in global star formation due
to gradual depletion of the gas.

\begin{figure}
\begin{center}
\includegraphics[width=84mm]{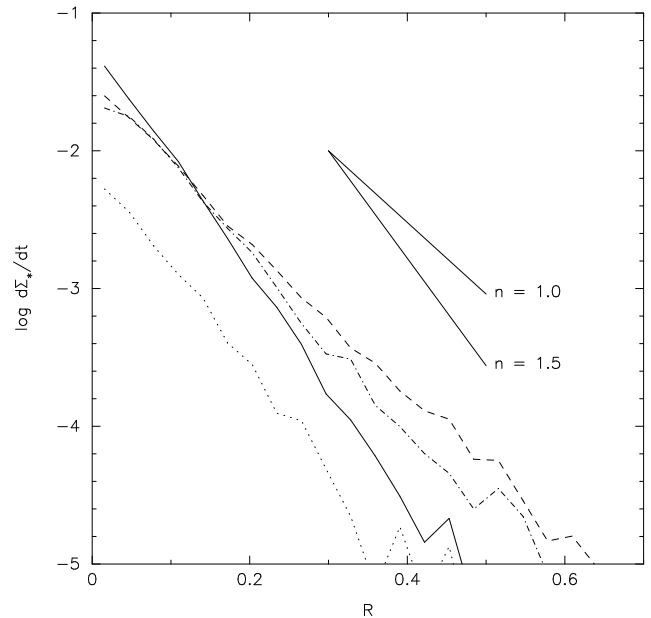}
\end{center}
\caption{Radial star formation profiles for isolated disk simulations.
Line types are identical to those in Fig.~\ref{fig06}.  The two
labeled lines correspond to Schmidt laws with the indicated values of
$n$.}
\label{fig07}
\end{figure}

Given the rather stable behavior of the isolated disk models, it's
reasonable to measure radial star formation profiles by counting stars
formed in concentric annuli.  Fig.~\ref{fig07} plots
$\dot{\Sigma}_{*}$ as a function of cylindrical radius.  Much as
expected, the density-dependent rule yields a good approximation to
the standard Schmidt law with index $1.5$.  Less predictably, the
shock-induced rules also approximate Schmidt-type laws, albeit with
somewhat shallower slopes.  This outcome must depend on the dynamical
structure of the disks; for a hypothetical and rather extreme example,
a bar-unstable disk would presumably yield different star formation
profiles before and after the bar formed.  A shock-induced model in
which the average shock strength is independent of $R$ should yield a
Schmidt-type law with an index of unity, regardless of the chosen
value of $m$.  Here, however, the shock-induced models with $m = 1.0$
appear to yield slightly steeper $\dot{\Sigma}_{*}$ profiles than does
the model with $m = 0.5$, hinting that the shocks are somewhat
stronger near the center of this disk.

\end{document}